# Experimental and theoretical determination of σ-bands on ("2√3×2√3") silicene grown on Ag(111)

W. Wang[*], W. Olovsson and R. I. G. Uhrberg

*Department of Physics, Chemistry, and Biology, Linköping University, S-581 83 Linköping, Sweden*

Silicene, the two-dimensional (2D) allotrope of silicon has very recently attracted a lot of attention. It has a structure that is similar to graphene and it is theoretically predicted to show the same kind of electronic properties which has put graphene into the focus of large research and development projects world-wide. In particular, a 2D structure made from Si is of high interest because of the application potential in Si-based electronic devices. However, so far there is not much known about the silicene band structure from experimental studies. A comprehensive study is here presented of the atomic and electronic structure of the silicene phase on Ag(111) denoted as (2√3×2√3)R30° in the literature. Low energy electron diffraction (LEED) shows an unconventional rotated ("2√3×2√3") pattern with a complicated set of split diffraction spots. Scanning tunneling microscopy (STM) results reveal a Ag(111) surface that is homogeneously covered by the ("2√3×2√3") silicene which exhibits an additional quasi-periodic long range ordered superstructure. The complex structure, revealed by STM, has been investigated in detail and we present a consistent picture of the silicene structure based on both STM and LEED. The homogeneous coverage by the ("2√3×2√3") silicene facilitated an angle-resolved photoelectron spectroscopy study which reveals a silicene band structure of unprecedented detail. Here, we report four silicene bands which are compared to calculated dispersions based on a relaxed (2√3×2√3) model. We find good qualitative agreement between the experimentally observed bands and calculated silicene bands of σ character.

PACS number: 68.37.Ef, 71.15.Mb, 73.20.-r, 79.60.-i

## I. INTRODUCTION

The two-dimensional (2D) honeycomb structure of carbon (C) atoms, i.e., graphene, has attracted great attention because of its outstanding electronic properties due to $sp_2$ hybridization. The electronic structure shows a steep linear dispersion near the Fermi energy ($E_F$) forming a so called Dirac cone at specific points in k-space. This part of the electronic structure resembles what is expected for massless fermions. Inspired by the exceptional properties of graphene, a lot of effort now goes into the search for other 2D $sp_2$ hybridized materials. As the nearest neighbor to C in group IV of the periodic table, Si is one of the most promising candidates. A 2D graphene like structure made from Si is of high interest because of the application potential in Si-based electronic devices. Such a 2D honeycomb structure of Si has been theoretically studied since the 1980's[1] and was named silicene in 2007 by Guzmán-Verri and Yan Voom[2]. Unlike graphene, free-standing silicene is calculated to be stable only in a slightly buckled geometry. The reason is that the s and p valence orbitals of Si are prone to form an $sp_3$ hybridization which leads to a tetrahedral, three dimensional, configuration rather than the planar $sp_2$ hybridization of a two dimensional structure. For a small buckling, calculations do predict an electronic structure with Dirac cones[3] at the $\bar{K}$-points of the silicene Brillouin zone. However, there is no convincing experimental evidence of the presence of Dirac fermions in monolayer silicene. An electron band with a linear dispersion was found by Vogt *et al.*[4] near a $\bar{K}$-point of silicene grown on Ag(111) and it was discussed as emission from a modified Dirac cone with a gap between π and π* bands. However, this idea has been questioned by several authors,[5-8] and the linear band is instead suggested to originate from 2D surface or interface states that appear when the silicene layer has formed. As suggested by Cahangirov *et al.*[9] the linear dispersion found in Ref. 4 can arise from new states due to strong hybridization between Si and Ag orbitals.

Silicene is mostly synthesized on Ag(111) substrates because of a low Si-Ag intermixing and a "matching" 3 to 4 ratio between the lattice constant of silicene and that of the Ag(111) surface.[4,5,10-15] Several modifications of silicene with different mixes of $sp_2$ and $sp_3$ hybridization and different orientations with respect to the Ag surface lattice have been observed. Depending on the orientation, the silicene layer shows specific reconstructions involving buckling and thus a deviation from an "ideal" $sp_2$ hybridization. The most commonly observed periodicities are (4×4) α=0°, (√13×√13) Type I α=±27°, (√13×√13) Type II α=±5.2°, (√7×√7) α=±19.1°, and (2√3×2√3) α=±10.9°, where α is the nominal angle between the silicene and the Ag(111) surface unit cells derived from atomic models[10]. The formation of the different reconstructions depends on substrate temperature, deposition rate and coverage. They all form within a small region of parameter space and samples generally exhibit more than one silicene reconstruction[16-19]. Angle-resolved photoelectron spectroscopy (ARPES) experiments have focused on the (4×4) phase since it is aligned with the (1×1) unit cell of Ag(111)[4,5,7]. However, to the best of our knowledge, there is no clear evidence, neither from real space, nor from reciprocal space studies of a (4×4) silicene surface without the coexistence of other phases. The difficulty to grow large domains of a single silicene phase has so far hampered investigations of the electronic structure of individual silicene phases by ARPES.

In this paper, we present a comprehensive study of a silicene layer that is known in the literature as (2√3×2√3)R30° silicene due to its approximate periodicity with respect to the Ag(111) surface lattice, as observed by LEED[10,20]. This structure is formed by silicene layers that are rotated by either +10.9° or -10.9° with respect to Ag(111). The presence of both orientations of the silicene layer was verified experimentally in our study (see supplementary information[21]). Structural information was obtained by STM and LEED, and the electronic structure was determined by ARPES. By comparing to the results of electronic band structure calculations, based on a (2√3×2√3) model of the silicene layer, we could unambiguously identify electronic bands of silicene (σ-bands) in the ARPES data.

## II. EXPERIMENTAL AND THEORETICAL DETAILS

Samples were prepared in-situ in ultrahigh vacuum (UHV) systems with base pressures in the $10^{-11}$ Torr range. The Ag(111) crystal was cleaned by repeated cycles of sputtering by $Ar^+$ ions (1 keV) and annealing at approximately 400 °C until a sharp (1×1) LEED pattern was obtained. Atomically resolved STM images, as well as the presence of a sharp Shockley surface state in ARPES, confirmed a high quality of the Ag(111) surface. About 1 monolayer (ML) of Si was deposit from a heated Si wafer piece at a rate of about 0.03 ML/min. The Ag(111) substrate was kept between 280 and 300 °C during deposition. STM images were recorded at room temperature using an Omicron variable temperature STM at Linköping University. LEED and ARPES data were obtained at the MAX-lab synchrotron radiation facility using the beam line I4 end station. Data were acquired at room temperature by a Phoibos 100 analyzer from Specs with a two-dimensional detector. The energy and angular resolutions were 50 meV and 0.3°, respectively. Angle integrated Si 2p core-level spectra were measured at room temperature. DFT calculations were performed using the WIEN2k package[22] employing the generalized gradient approximation (GGA) for exchange and correlation. The periodic slab model includes nine layers of Ag atoms, one layer of silicene and a vacuum layer of 15 Å. The positons of the atoms were fully relaxed using the projector augmented wave method (PAW) Vienna *ab initio* simulation package (VASP) code[23] except for the bottom six Ag layers. The energy cutoff of the plane-wave basis set was 250 eV, and the k-point mesh was 4×4×1.

## III. RESULTS AND DISCUSSION

Figure 1(a) is a large-scale ~90×90 nm² topographic filled-state image showing an overview of the (2√3×2√3) silicene layer. STM images obtained at different places on the surface verified the homogeneous coverage of the Ag(111) surface by (2√3×2√3) silicene. It should be noted that there are two differently oriented domains, differing by a rotation angle of ~8°. The 8° difference implies that there are two variants of the (2√3×2√3) reconstruction, one rotated by 26° and the other by 34° with respect to Ag(111). From here on, we use the short notation "2√3", where the quotation marks indicate that the real silicene layer deviates from an ideal (2√3×2√3) periodicity.

A close-up image of a 42×42 nm² area (Fig. 1(b)) shows a clear quasi-periodic long-range order in the form of brighter parts arranged in a hexagonal fashion. The orientation of this hexagonal structure is rotated by ~9° clockwise relative to the Ag(111) surface, as illustrated by the white and green dashed lines. This phase has been reported by Feng *et al.*[24], which they described as a (2√3×2√3) reconstruction with an obvious moiré pattern, due to the quasi-periodic brighter parts observed by STM, see Fig. 1(b). This long-range pattern is actually not a consequence of moiré interference but a consequence of a mismatch of about 2 % in the lattice constants of silicene and Ag(111)[10,25], resulting in ideal and distorted honeycomb rings in bright and dark parts respectively[26]. Figure 1(c) is a detailed STM image (3.8×3.8 nm²) of Fig. 1(b). A fast Fourier transform (FFT) of the STM image in Fig. 1(a), results in two sets of Fourier components, each set forming a hexagonal pattern, see Fig. 1(d). The split into two hexagonal sets of components is just a consequence of the presence of two "2√3" domains differing by ~8°. This FFT map can be directly compared to the LEED pattern in Fig. 1(e) to gain important information about this complex surface structure. The strong Fourier components within the green and blue circles of Fig. 1(d) correspond to the two brightest spots within the circles of the LEED pattern. Thus, these LEED spots originate from two ±4° rotated "2√3" domains explaining the major deviation from an ideal (2√3×2√3) diffraction pattern. Figure 1(e) also shows two weaker spots within the circles which originate from the moiré-like hexagonal structure. One set of spots is rotated by +25° with respect to the -4° domain of the "2√3" phase and the other set of spots is rotated by -25° with respect to the +4° domain. The periodicity of the moiré like pattern is approximately 3.7 times larger than that of the "2√3" structure. These assignments have been derived from a detailed FFT analysis of single domain STM images and experimental LEED patterns, see the supplementary information[21].

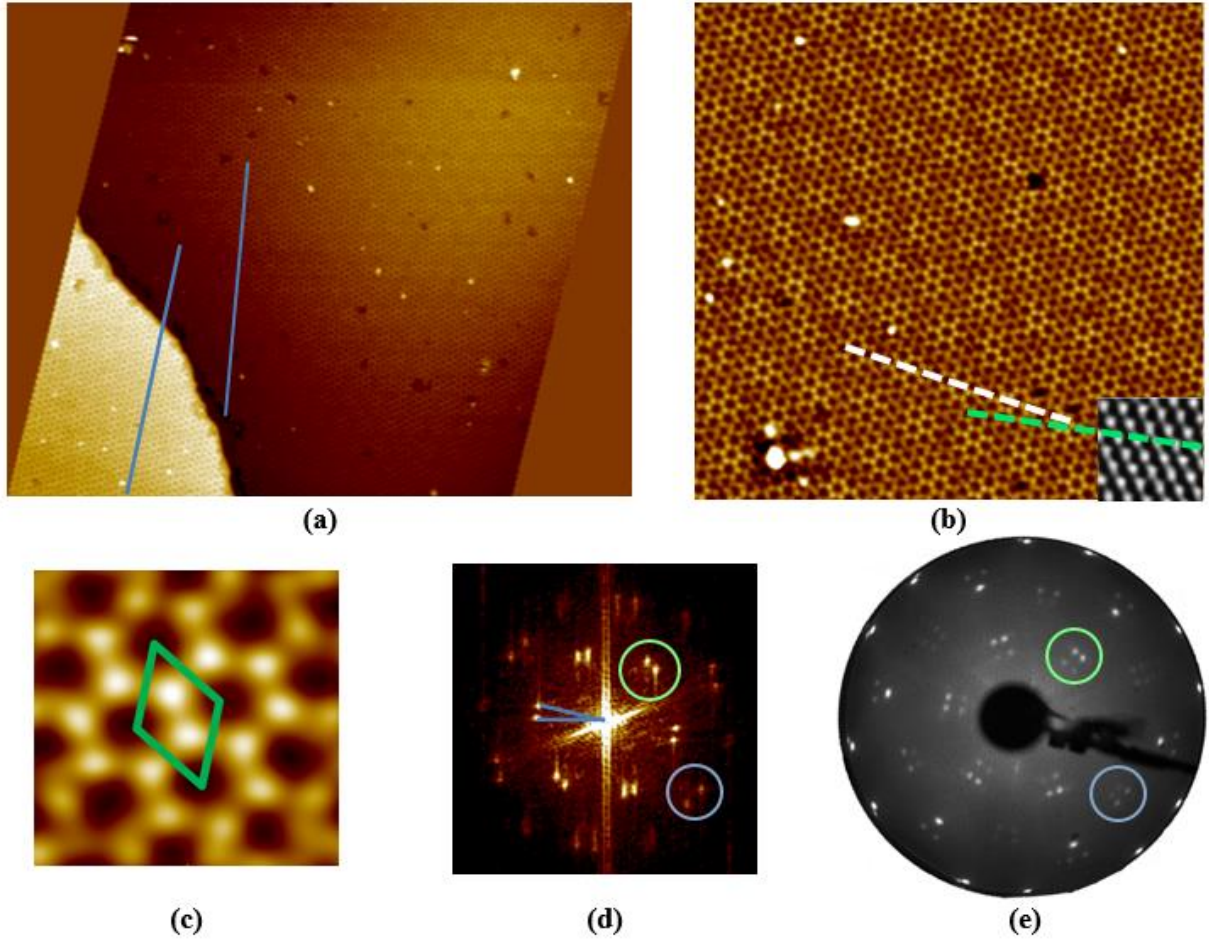

FIG. 1. (Color on line) (a) Large scale filled-state STM image, ~90×90 nm$^2$, of "2√3" silicene on two terraces of the Ag(111) substrate. The blue lines indicate the orientations of the two domains which differ by ~8°. The image has been compensated for thermal drift. (b) Atomically resolved filled state STM image of a 42×42 nm$^2$ area showing honeycomb rings and a long range, quasi-periodic, hexagonal pattern. The white dashed line is a guide to the eye of the orientation of this quasi-periodic pattern. The inset is a 1.8×1.8 nm$^2$ filled state STM image of the clean Ag(111) surface. The green dashed line indicates the orientation of the Ag(111) surface unit cell. The angle between the green and white dashed lines is about 9°. (c), Atomically resolved STM image of a 3.8×3.8 nm$^2$ area showing the well-ordered honeycomb rings within the brighter parts of the STM image in (b). The green diamond depicts a "2√3" unit cell. All STM images were recorded at room temperature in constant current mode with a tunneling current of 200 pA and a sample bias of -1.2 V in (a), -1.5 V in (b) and (c), and -0.1 V for the clean Ag(111) (inset in (b)). (d) Fast Fourier transform (FFT) of the STM image in (a) showing two sets of Fourier components forming two hexagons with a rotational difference of ~8° (±4° relative to the Ag(111) lattice) as indicated by the blue lines. (e) LEED pattern obtained from "2√3" silicene at an electron energy of 40 eV. The two bright spots inside each circle are due to diffraction from the two "2√3" domains. These diffraction spots match the split Fourier components encircled in (d). The two weaker LEED spots inside the circles originate from the long range, quasi-periodic, structure. A detailed analysis of STM, FFT and LEED results is available in the supplementary information[21].

The homogeneous coverage of the Ag(111) surface by the "2√3" silicene layer opens up for an unambiguous ARPES study of the electronic structure. Initially, the clean Ag(111) substrate was studied in order to establish its contribution to the ARPES data. Figure 2(a) (left) shows bulk emission (B) and some emission intensity with a symmetric dispersion around the $\bar{M}$ point of the Ag (1×1) surface Brillouin zone (SBZ) using a photon energy of 19 eV. After the formation of the "2√3" silicene, there is no change in the bulk emission, while there is a clear change of the emission symmetric around the $\bar{M}$ point which now exhibits an energy maximum well below $E_F$ at 0.4 eV (middle). A comparison of spectra obtained at two different photon energies shows that the bulk emission B crosses $E_F$ at different k-values, (1.11Å$^{-1}$ when the photon energy is 19 eV (middle) and 1.16 Å$^{-1}$ in the 26 eV data (right)). The band, $S_{Ag}$, does not change the shape and the energy maximum remains at the same position, which confirm its 2D character. In a recent paper[8], a calculation was presented indicating that the emission here labeled $S_{Ag}$ is actually due to hybridization between Si and Ag orbitals. A similar band is also observed by ARPES for the $Ag_2Sn$ surface alloy on Ag(111)[27], which supports the interpretation that $S_{Ag}$ is not specifically related to the silicene layer, but instead stems from a modification of the electronic structure of the outermost Ag layer.

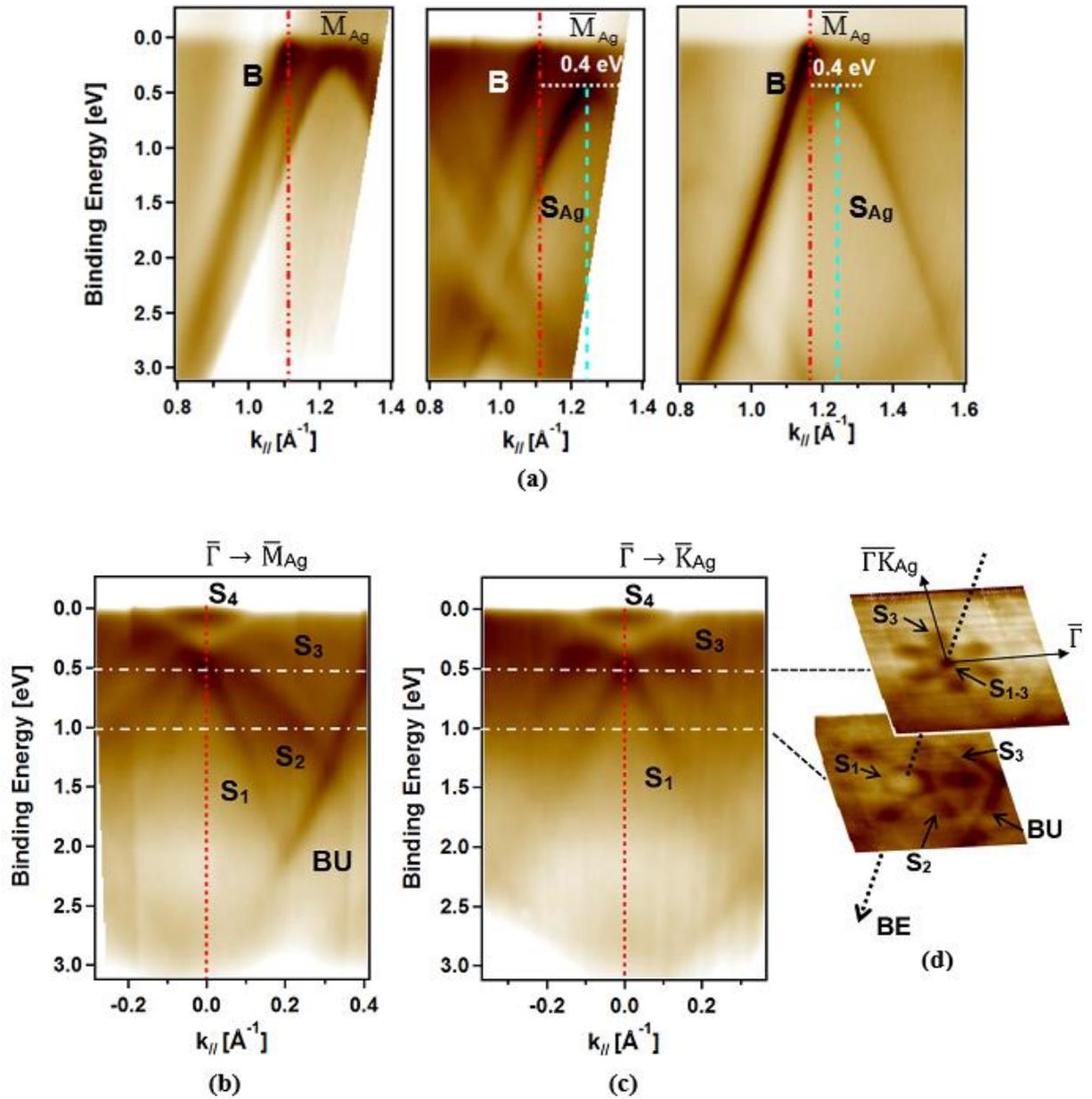

FIG. 2. (Color online) (a) Valence band dispersions of i) clean Ag(111) using a photon energy of 19 eV (left), ii) "2√3" silicene on Ag(111) using 19 eV (middle) and 26 eV photons (right). The data were obtained around $\bar{M}$ in the $\bar{\Gamma}\bar{M}$ direction of the 1×1 SBZ of Ag(111). The $S_{Ag}$ band appears after the formation of silicene. The dispersion does not change with photon energy, compare middle and right panels, but the bulk related emission B does. This verifies the 2D character of $S_{Ag}$. (b) and (c) Valence band dispersions of "2√3" silicene on Ag(111) obtained using a photon energy of 19 eV. The bands were mapped along the $\bar{\Gamma}\bar{M}$ and $\bar{\Gamma}\bar{K}$ lines of the Ag (1×1) SBZ, respectively. Three dispersive silicene bands, $S_1$, $S_2$ and $S_3$, are observed around $\bar{\Gamma}$. A dispersive feature (BU) due to umklapp scattering of the bulk Ag sp emission (B) is also observed and the position fits with scattering by a 2√3 reciprocal lattice vector. A fourth silicene related feature, $S_4$, is observed as an almost flat band at $E_F$ close to $\bar{\Gamma}$. (d) Constant energy contours obtained at 0.5 and 1.0 eV below $E_F$. The $S_1$, $S_2$ and $S_3$ bands seem to be degenerate at $\bar{\Gamma}$ at 0.5 eV leading to a spot at the center of the 0.5 eV contour. The six lobes stretching out from $\bar{\Gamma}$ originate from the $S_3$ band. In the 1.0 eV contour plot, the $S_1$ band gives rise to the inner hexagon, six lobes are generated by emission from $S_2$, while the $S_3$ emission results in the outer weak hexagon. The contour due to the BU emission is also indicated.

The experimental 2D band structure of "2√3" silicene is presented along the $\bar{\Gamma}\bar{M}$ and $\bar{\Gamma}\bar{K}$ lines of the Ag (1×1) SBZ in Figs. 2(b) and 2(c), respectively. Two downward dispersing bands, $S_1$ and $S_2$, are clearly observed in the $\bar{\Gamma}\bar{M}$ direction, while a third band, $S_3$, with an upward dispersion starting from $\bar{\Gamma}$ can be better seen along $\bar{\Gamma}\bar{K}$ where the upward dispersion is evident changing to a downward dispersion further out in the SBZ. Along $\bar{\Gamma}\bar{K}$, $S_2$ cannot be detected while $S_1$ appears with an intensity and dispersion similar to the results along $\bar{\Gamma}\bar{M}$. The dispersive feature, BU, in Fig. 2(b) is due to umklapp scattered Ag bulk emission appearing in the outer parts of the Ag (1×1) SBZ, see feature B in Fig. 2(a). In addition to these three dispersive bands, there is electron emission, $S_4$, close to $E_F$ at normal emission. On samples which are only partly covered by silicene, we observed the Shockley surface state of clean Ag(111) near normal emission. However, it is clearly separated from the $S_4$ emission and we therefore also assign $S_4$ to the silicene layer. Constant energy contours are shown in Fig. 2(d) at binding energies of 0.5 and 1.0 eV. The contour of $S_1$ is a hexagon with the same orientation as the hexagon formed by the Ag(111) (1×1) SBZ, while the $S_2$ and $S_3$ bands give rise to a six-fold pattern of lobes, pointing along $\overline{\Gamma M}$ and $\overline{\Gamma K}$, respectively.

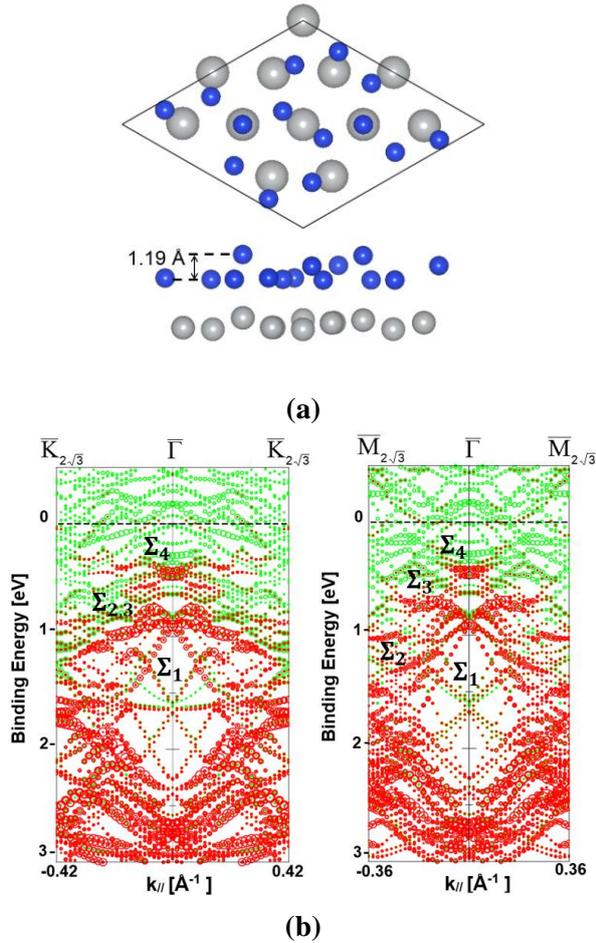

FIG. 3. (Color online) (a) Top and side views of a unit cell of the relaxed model of (2√3×2√3) silicene on Ag(111). Only the Si atoms (blue) and the first layer Ag atoms (grey) are included. The Si atoms on top of the silver atoms give rise to the honeycomb pattern observed by STM. The slab used in the band structure calculations is composed of one buckled Si layer, 3 relaxed, slightly buckled, Ag layers and 6 bulk Ag layers. (b) Calculated band structure for (2√3×2√3) silicene on Ag(111) where only contributions from the Si atoms are plotted. The red and green circles represent the $p_xp_y$ and $p_z$ states of the Si atoms, respectively. The bigger the circles are, the larger the contribution is from the Si atoms to the wave functions at the various k-points. The band structure is plotted in the 2√3×2√3 SBZ. Note that the $\overline{\Gamma K}_{2\sqrt{3}}$ and $\overline{\Gamma M}_{2\sqrt{3}}$ lines correspond to $\overline{\Gamma M}_{Ag}$ and $\overline{\Gamma K}_{Ag}$ in Figs. 2(b) and 2(c) respectively. Several of the calculated bands, labeled $\Sigma_1$-$\Sigma_4$, show qualitative agreement with the experimental bands $S_1$-$S_4$ in Figs. 2(b) and 2(c). The calculated bands below a binding energy of about 2 eV were not observed experimentally.

In order to further analyze the $S_1$-$S_4$ bands, we performed first-principles density functional theory (DFT) calculations using the full-potential linearized augmented plane wave (FPLAPW) WIEN2k simulation package[22]. We model the experimental "2√3" surface by the structure shown in Fig. 3(a). This model should quite accurately represent the structure of the brighter parts in the STM images (see Fig. 1). The so called (2√3 × 2√3) type II model in Ref. 18 and (√7×√7) silicene on (2√3×2√3) Ag(111) model in Ref. 25 presented similar atomic structures as our model by showing two topmost Si atoms in the 2√3 unit cell sitting on top of Ag atoms. The height difference between highest and lowest Si atoms is 1.0 Å[18] and 1.19Å[25] respectively, which agree nicely with our value of 1.19Å. The silicon atoms that are located on top of silver atoms form a honeycomb pattern. Between the bright parts, the honeycombs show some distortions which are not included in the model. Calculated bands along the $\overline{\Gamma K}_{2\sqrt{3}}$ and $\overline{\Gamma M}_{2\sqrt{3}}$ symmetry lines of the 2√3×2√3 SBZ are shown in Fig. 3(b). Red and green circles represent $p_xp_y$ (σ) and $p_z$ (π) bands of the silicene layer, respectively. The bigger the circles are, the larger the contribution is from the Si atoms to the wave functions at the various k-points. Comparing the experimental and calculated bands, one finds striking qualitative similarities in the binding energy range from 0 to ~1.5 eV. i) The steep $\Sigma_1$ band in the calculation reproduces the dispersion of the $S_1$ band. The $\Sigma_1$ energy at $\overline{\Gamma}$ is 0.9 eV, while the experimental value is 0.5 eV. ii) The upward dispersive "bands" labeled $\Sigma_{2,3}$ and $\Sigma_3$ close to $\overline{\Gamma}$ agree qualitatively with the experimental $S_3$ band. It should be noted that the particular shape shown by $\Sigma_{2,3}$ and $\Sigma_3$ has not been reported in any published band structure calculation of silicene on Ag(111). Thus, the initial upward dispersion observed experimentally, which is reproduced theoretically, appears to be a consequence of the particular buckling of the silicene layer in this case. Between $\Sigma_1$ and $\Sigma_3$ there are some dispersive bands, indicated by $\Sigma_2$, which fit qualitatively with the experimental $S_2$ band. Confined to a small region around $\overline{\Gamma}$, there are silicene states $\Sigma_4$, at 0.5 eV, i.e., 0.4 eV higher than the $\Sigma_1$ maximum. These theoretical findings fit nicely with the characteristics of $S_4$ which we therefore assign to Si $p_xp_y$ states. To summarize, we find that the model calculation reproduces the shapes of the $S_1$, $S_3$ and $S_4$ bands quite well and, to a somewhat lesser extent, also the $S_2$ band. Quantitatively, the calculation locates the bands ~0.4 eV lower in energy. A reason for the difference can be that in the present calculations self-energy effects are not included to obtain the quasiparticle band structure[28]. Even though the (2√3×2√3) model used for calculation does not exactly represent the real "2√3" silicene structure, we conclude that the experimental bands observed by ARPES originate from the silicene layer. The comparison with theory identifies them as bands of mainly σ character. This is the first report of electron bands that can be unambiguously assigned to silicene.

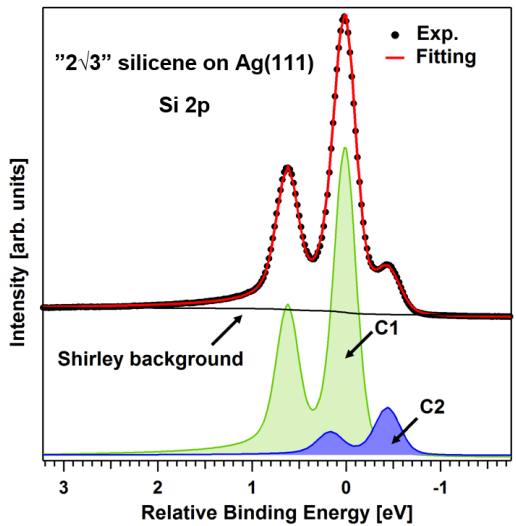

FIG. 4. (Color online) Si 2p core-level spectrum obtained at a photon energy of 135 eV in normal emission. The dots are the experimental data and the fitting curve is the sum of two spin-orbit split components (C1 and C2) and a Shirley background. The spectrum is dominated by C1. The C2 component, which has ~15% of total intensity, is shifted from C1 by -0.46 eV. Fitting parameters: Spin-orbit split: 0.61 eV; Branching ratio: 0.46; Gaussian width: ~220 meV (C1) and ~260 meV (C2); Lorentzian width: 80 meV (C1 and C2). The asymmetry parameter of the Doniach–Šunjić line profile is 0.055.

Angle-integrated Si 2p core-level spectra were also measured in order to gain further information about the atomic and electronic structures. Figure 4 shows a Si 2p spectrum measured from "$2\sqrt{3}$" silicene using a photon energy of 135 eV. The spectrum consists of two components, C1 and C2. The intensity of C2 corresponds to ~15 % of the total intensity. It is interesting to compare this result to the expectations based on the ($2\sqrt{3}\times2\sqrt{3}$) model in Fig. 3(a). The two Si atoms, located on top of Ag atoms, constitute 14 % of the 14 Si atoms in the unit cell. These atoms are the ones observed by STM both in the bright and dark parts, indicating that they are markedly different from the other silicon atoms. It is therefore natural to assign the C2 component of the Si 2p spectrum to these atoms. Although, the real "$2\sqrt{3}$" silicene layer shows deviations from the atomic structure of the ($2\sqrt{3}\times2\sqrt{3}$) model, we conclude that the Si 2p core-level data are consistent with the model used in this study.

## IV. SUMMARY

By analyzing LEED and STM results, we have obtained a detailed structure determination of "$2\sqrt{3}$" silicene on Ag(111). Due to strain, the silicene layer exhibits a quasi-periodic long-range modulation resulting in a hexagonal like superstructure with a lattice constant ~3.7 times larger than that of a $2\sqrt{3}$ unit cell. The homogeneous coverage of the Ag(111) surface by this type of silicene made it meaningful to perform an ARPES study. As the main result, we have reported four silicene bands, $S_1$-$S_4$. By comparing to a band structure calculation on a relaxed ($2\sqrt{3}\times2\sqrt{3}$) model we could unambiguously assign $S_1$-$S_3$ to σ-bands of the silicene sheet.

## ACKNOWLEDGMENTS

Technical support from Dr. Johan Adell, Dr. Craig Polley and Dr. T. Balasubramanian at MAX-lab is gratefully acknowledged. Financial support was provided by the Swedish Research Council (Contracts No. 621-2010-3746, 621-2014-4764 and 621-2011-4426) and by the Linköping Linnaeus Initiative for Novel Functional Materials supported by the Swedish Research Council (Contract No. 2008-6582). The calculations were carried out at the National Supercomputer Centre (NSC), supported by the Swedish National Infrastructure for Computing (SNIC).

[1] M.T. Yin and M.L. Cohen, Phys. Rev. B **29**, 6996 (1984)
[2] G.G. Guzmán-Verri and L.C. Lew Yan Voon, Phys. Rev. B **76**, 075131 (2007)
[3] S. Cahangirov, M. Topsakal, E. Aktürk, H. Sahin and S. Ciraci, Phys. Rev. Lett. **102**, 236804 (2009)
[4] P. Vogt *et al.*, Phys. Rev. Lett. **108**, 155501 (2012)
[5] D. Tsoutsou, E. Xenogiannopoulou, E. Golias, P. Tsipas, and A. Dimoulas, Appl. Phys. Lett. **103**, 231604 (2013)
[6] C.-L. Lin *et al.*, Phys. Rev. Lett. **110**, 076801 (2013)
[7] S.K. Mahatha *et al.*, Phys. Rev. B **89**, 201416(R) (2014)
[8] Y.-P. Wang and H.-P. Cheng, Phys. Rev. B **87**, 245430 (2013)
[9] S. Cahangirov *et al.*, Phys. Rev. B **88,** 035432 (2013)
[10] H. Jamgotchian *et al.*, J. Phys.: Condens. Matter **24**, 172001 (2012)
[11] Z.-X. Gao, S. Furuya, J. Iwata and A. Oshiyama, J. Phys. Soc. Jpn. **82**, 063714 (2013)
[12] L. Chen *et al.*, Phys. Rev. Lett. **110**, 085504 (2013)
[13] P. Vogt *et al.*, Appl. Phys. Lett. **104**, 021602 (2014)
[14] J. Avila *et al.*, J. Phys.: Condens. Matter **25**, 262001 (2013)
[15] S. Cahangirov *et al.*, Phys. Rev. B **90**, 035448 (2014)
[16] P. Moras, T. O. Mentes, P. M. Sheverdyaeva, A. Locatelli and C. Carbone, J. Phys.: Condens. Matter **26**, 185001 (2014)
[17] C. -L. Lin *et al.*, Appl. Phys. Express **5**, 045802 (2012)
[18] H. Enriques *et al.*, J. Phys. Condens. Matters **24**, 314211 (2012)
[19] M. R. Tchalala *et al.*, Appl. Surf. Sci. **303**, 61 (2014).
[20] A. Acun, B. Poelsema, H. J. W. Zandvliet and R. van Gastel, Appl. Phys. Lett. **103**, 263119 (2013)
[21] See Supplemental Material at [URL] for detailed analysis of LEED patterns and FFTs of STM images.
[22] P. Blaha, K. Schwarz, G. K. H. Madsen, D. Kvasnicka and J. Luitz, *WIEN2k: An Augmented Plane Wave+Local Orbitals Program for Calculating Crystal*

*Properties* (Karlheinz Schwarz, Technische Universität, Wien, Austria, 2001).

[23] G. Kresse and D. Joubert, Phys. Rev. B **59**, 1758 (1999)

[24] B. Feng *et al*., Nano Lett. **12**, 3507 (2012)

[25] J. Gao and J. Zhao, Sci. Rep. **2**, 861 (2012)

[26] H. Jamgotchian *et al*., J. Phys.: Condens. Matter **27**, 395002 (2015)

[27] J. R. Osiecki and R. I. G. Uhrberg, Phys. Rev. B **87**, 075441 (2013)

[28] F. Aryasetiawan and O. Gunnarsson, Rep. Prog. Phys. **61**, 237 (1998)

Supplementary information:

# Experimental and theoretical determination of σ-bands on ("2√3×2√3") silicene grown on Ag(111)

W. Wang[*], W. Olovsson and R. I. G. Uhrberg, Department of Physics, Chemistry, and Biology, Linköping University, S-581 83 Linköping, Sweden

* Corresponding author: W. Wang, weiwa49@ifm.liu.se

In this supplementary section, we present a detailed analysis of the complex low energy electron diffraction (LEED) pattern of the so called (2√3×2√3) reconstructed silicene layer on Ag(111). The published LEED patterns associated with this type of silicene show some differences. Jamgotchian et al.[1] present a pattern with just 2√3 spots, while in the study by Acun et al.[2] the LEED pattern shows some split spots. The more detailed pattern in the latter study indicates that the (2√3×2√3) periodicity is just approximate and we therefore use the notation "2√3" from here and onward. We have found that the appearance of the weaker split spots is strongly dependent on the quality of the surface. Figure S1 shows two LEED patterns obtained at electron energies of 75 and 40 eV, which show a wealth of stronger and weaker diffraction spots. The pattern can be characterized as showing various clusters of spots around 2√3 positions. Hence, the LEED pattern may look like a (2√3×2√3) pattern if the splitting of the spots is not resolved.

By a combination of scanning tunneling microscopy (STM) images, fast Fourier transforms (FFTs) of STM images, and schematic LEED patterns, we have performed a detailed analysis of the origins of the diffraction spots observed in Fig. S1. The analysis is described in detail by the captions of Figs. S1-S4. The result is summarized at the end of this supplementary section.

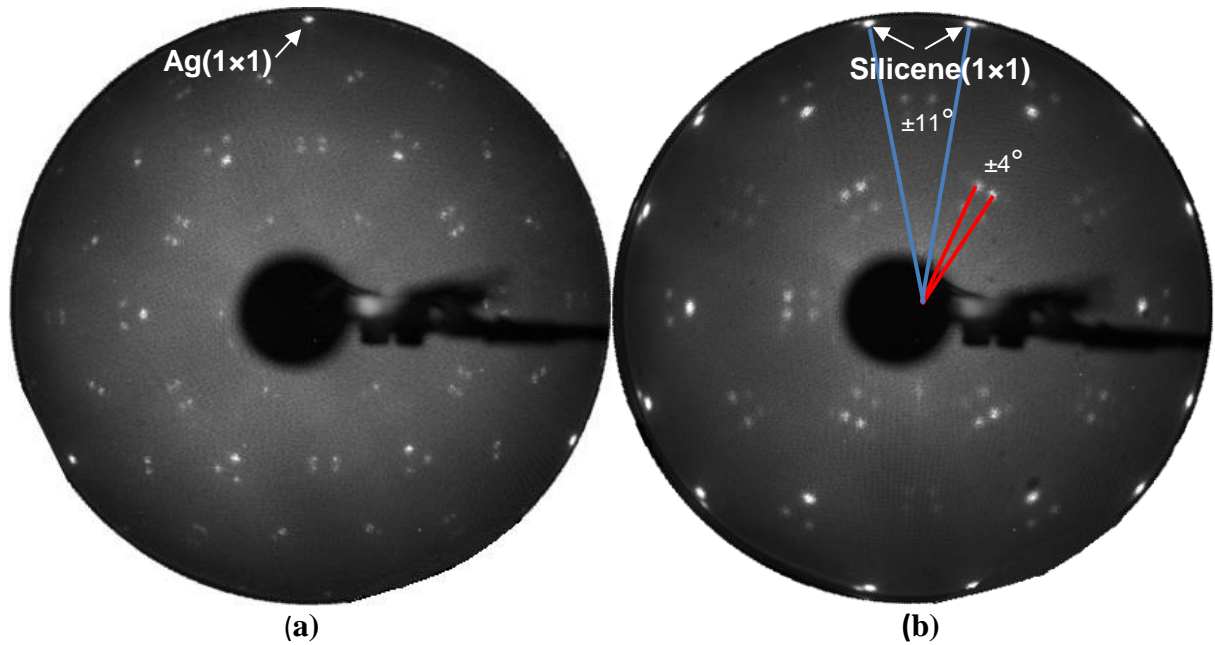

FIG. S1. (Color online) (a), LEED pattern (75 eV) of silicene on Ag (111) exhibiting a complex diffraction pattern which has been associated with a (2√3×2√3)R30° periodicity in the literature. However, instead of a single spot at each 2√3 position, the LEED pattern shows a set of spots with a systematic variation of the local configuration near 2√3 positions. We here use the notation "2√3" for the complex LEED pattern and the corresponding real space structure. (b) LEED pattern (40 eV) revealing more details. The blue lines indicate 1×1 spots belonging to different silicene domains rotated by ±11° *with respect to the Ag(111) substrate* (±10.9° in theoretical model[1]). *The red lines point at* spots from two "2√3" domains rotated by ±4° *relative to an ideal* (2√3×2√3) pattern based on *Ag(111).*

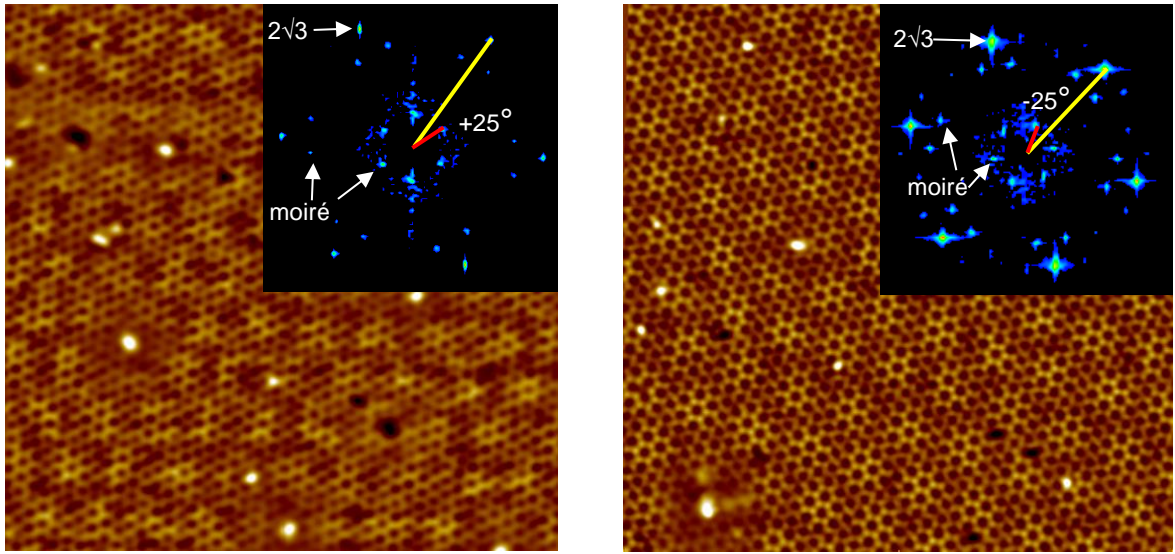

FIG. S2. (Color online) (a) STM image of the -4° oriented "2√3" domain, ~38×38 nm². (b) STM image of the other domain, rotated by +4° with respect to an ideal 2√3 pattern, ~38×38 nm². Both images show a similar hexagonal moiré-like pattern that is quasi periodic, but oriented differently. The insets in (a) and (b) are FFTs of the STM images of the two domains. The yellow lines indicate the orientation of a primitive vector for the "2√3" periodicity for each domain. As can be derived from the FFT maps, there is an angle difference of 8°, which corresponds to the LEED observation of diffraction from the two ±4° domains shown in Fig. S1. The center of the FFT maps and each "2√3" FFT component are surrounded by a smaller hexagon which is formed by the FFT components of the moiré-like structure. These hexagons are about 3.7 times smaller than the "2√3" hexagon and the orientation, as indicated by the red line, is +25° in (a) and -25° in (b) relative to the "2√3" orientation shown by the yellow line in each case. The STM images were obtained at room temperature using a bias of -1.5 V, and a tunneling current of 200 pA. The STM image in (a) is a zoom-in of a large scale image resulting in a lower resolution compared to the STM image in (b).

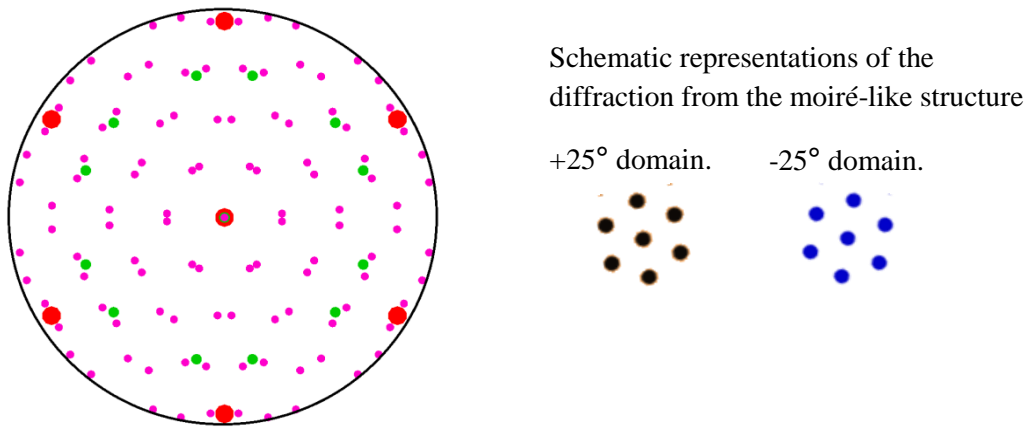

FIG. S3. (Color online) Schematic LEED pattern based on the experimental LEED and STM results of Figs. S1 and S2. Ag (1×1) spots are shown by the red solid circles. The two ±11° domains of silicene are shown by green spots using the silicene lattice constant. The pink spots correspond to the ±4° orientations of the "2√3" reconstruction. To the right of the schematic LEED pattern, two hexagonal patterns are drawn showing the orientation of the expected diffraction related to the moiré-like structural modification. These patterns are added to the corresponding "2√3" spots and drawn to scale in Fig. S4.

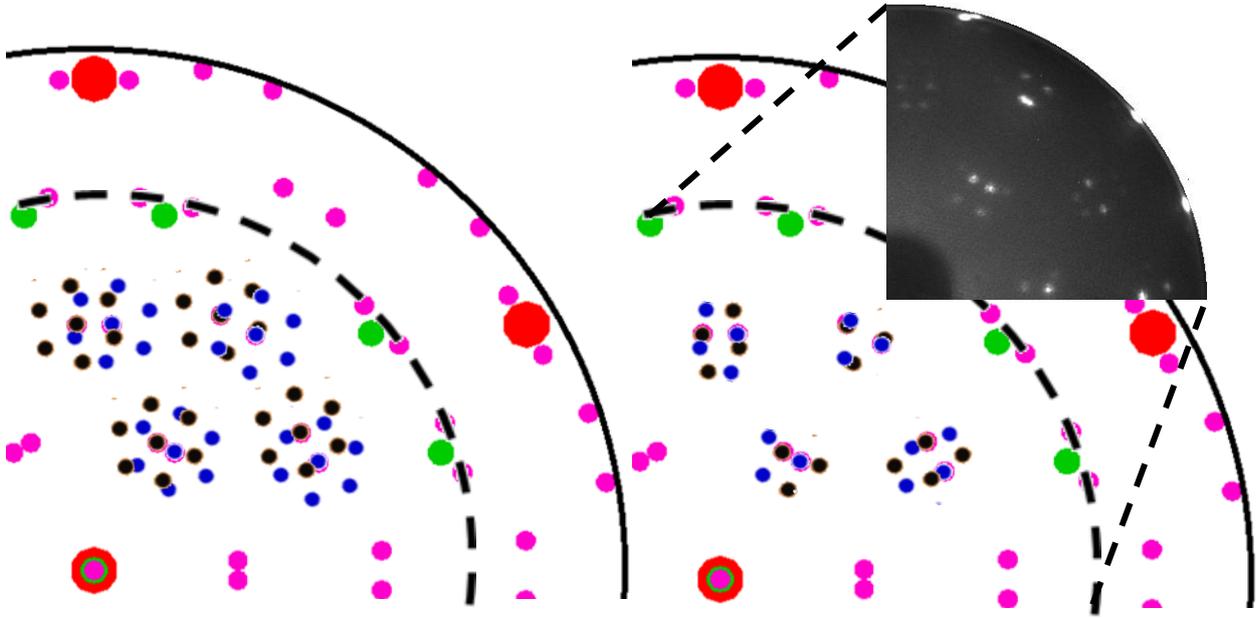

FIG. S4. (Color online) The simulated hexagons of the moiré-like pattern have been added to the respective LEED spots of the rotated "2√3" domains within the quadrant indicated by the dashed lines. The experimental LEED pattern does not show all the spots of the schematic pattern, but the ones that appear all fit with a subset of the schematic spots. In the right part of the figure, the spots of the schematic pattern which are observed by LEED have been singled out.

From the above analysis, we conclude that the LEED and STM results provide a consistent picture of the "2√3" structure. Silicene sheets of two orientations are present on the surface. The 1×1 unit cell of these sheets are rotated by ±11° with respect to the 1×1 unit cell of Ag(111). These silicene sheets are buckled forming quasi periodic structures with a "2√3" periodicity rotated by ±4°, respectively. The ordered regions form a moiré-like hexagonal pattern, observed by STM, which is rotated by approximately ±25° with respect to the "2√3" unit cell, or in another words, -9° (26° + 25°=51° which is equivalent to -9° for the six-fold symmetric structure) and 9° (34° - 25°=9°) relative to Ag (1×1). The complex sets of diffraction spots observed in LEED near (2√3×2√3) positions originate from the long range quasi periodic modulation of the silicene.

**References**

1. H. Jamgotchian *et al*., J. Phys.: Condens. Matter **24**, 172001 (2012)

2. A. Acun, B. Poelsema, H. J. W. Zandvliet and R. van Gastel, Appl. Phys. Lett. **103**, 263119 (2013)